\documentclass[review]{elsarticle}
\usepackage{hyperref,amssymb,graphicx,booktabs}

\journal{Nuclear Instruments and Methods in Physics Research A}

\newcommand{\muo}{ \mu_{\scriptscriptstyle 0}}

\begin{document}

\begin{frontmatter}

\title{Sensitivity of Fields Generated within Magnetically Shielded
  Volumes to Changes in Magnetic Permeability}

\author[manitoba]{T.~Andalib}
\author[winnipeg,manitoba]{J.W.~Martin\corref{mycorrespondingauthor}}
\cortext[mycorrespondingauthor]{Corresponding author}
\ead{j.martin@uwinnipeg.ca}
\author[winnipeg,manitoba]{C.P.~Bidinosti}
\author[winnipeg,manitoba]{R.R.~Mammei}
\author[winnipeg,manitoba]{B.~Jamieson}
\author[manitoba]{M.~Lang}
\author[triumf]{T.~Kikawa}

\address[winnipeg]{Physics Department, The University of Winnipeg, 515 Portage Avenue, Winnipeg, MB, R3B 2E9, Canada}
\address[manitoba]{Department of Physics and Astronomy, University of Manitoba, Winnipeg, MB R3T 2N2, Canada}
\address[triumf]{TRIUMF, 4004 Wesbrook Mall, Vancouver, BC V6T 2A3, Canada}

\begin{abstract}
Future experiments seeking to measure the neutron electric dipole
moment (nEDM) require stable and homogeneous magnetic fields.
Normally these experiments use a coil internal to a passively
magnetically shielded volume to generate the magnetic field.  The
stability of the magnetic field generated by the coil within the
magnetically shielded volume may be influenced by a number of factors.
The factor studied here is the dependence of the internally generated
field on the magnetic permeability $\mu$ of the shield material.  We
provide measurements of the temperature-dependence of the permeability
of the material used in a set of prototype magnetic shields, using
experimental parameters nearer to those of nEDM experiments than
previously reported in the literature.  Our measurements imply a range
of $\frac{1}{\mu}\frac{d\mu}{dT}$ from 0-2.7\%/K. Assuming typical
nEDM experiment coil and shield parameters gives
$\frac{\mu}{B_0}\frac{dB_0}{d\mu}=0.01$, resulting in a temperature
dependence of the magnetic field in a typical nEDM experiment of
$\frac{dB_0}{dT}=0-270$~pT/K for $B_0=1~\mu$T.  The results are useful
for estimating the necessary level of temperature control in nEDM
experiments.
\end{abstract}

\begin{keyword}
Magnetic Shielding \sep Neutron Electric Dipole Moment \sep Magnetic Field Stability
\end{keyword}

\end{frontmatter}


\section{Introduction}

The next generation of neutron electric dipole moment (nEDM)
experiments aim to measure the nEDM $d_n$ with proposed precision
$\delta d_n\lesssim
10^{-27}~e\cdot$cm~\cite{bib:nedm2,bib:nedm2.5,bib:nedm3,bib:nedm3.5,bib:nedm5,bib:nedm6,bib:nedm6.5,bib:nedmtriumf}.
In the previous best experiment \cite{bib:baker,bib:pendlebury} which
discovered $d_n<3.0\times 10^{-26}~e\cdot$cm (90\% C.L), effects
related to magnetic field homogeneity and instability were found to
dominate the systematic error.  A detailed understanding of passive
and active magnetic shielding, magnetic field generation within
shielded volumes, and precision magnetometry is expected to be crucial
to achieve the systematic error goals for the next generation of
experiments.  Much of the research and development efforts for these
experiments are focused on careful design and testing of various
magnetic shield geometries with precision
magnetometers~\cite{bib:brys,bib:afach,bib:fierlingerroom,bib:sturmthesis,bib:patton}.

In nEDM experiments, the spin-precession frequency $\nu$ of neutrons
placed in static magnetic $B_0$ and electric $E$ fields is measured.
The measured frequencies for parallel $\nu_+$ and antiparallel $\nu_-$
relative orientations of the fields is sensitive to the neutron
electric dipole moment $d_n$
\begin{equation}
h\nu_\pm=2\mu_nB_0\pm 2d_nE
\end{equation}
where $\mu_n$ is the magnetic moment of the neutron.

A problem in these experiments is that if the magnetic field $B_0$
drifts over the course of the measurement period, it degrades the
statistical precision with which $d_n$ can be determined.  If the
magnetic field over one measurement cycle is determined to $\delta
B_0=10$~fT, it implies an additional statistical error of $\delta
d_n\sim 10^{-26}~e\cdot$cm (assuming an electric field of $E=10$~kV/cm
which is reasonable for a neutron EDM experiment).  Over 100 days of
averaging, this would make a $\delta d_n\sim 10^{-27}~e\cdot$cm
measurement possible.  Unfortunately the magnetic field in the
experiment is never stable to this level.  For this reason,
experiments use a comagnetometer and/or surrounding atomic
magnetometers to measure and correct the magnetic field to this
level~\cite{bib:baker,bib:brys,bib:afach}.  Drifts of 1-10~pT in $B_0$
may be corrected using the comagnetometer technique, setting a goal
magnetic stability for the $B_0$ field generation system in a typical
nEDM experiment.

In such experiments, typically $B_0=1~\mu$T is used to provide the
quantization axis for the ultracold neutrons.  The $B_0$ magnetic
field generation system typically includes a coil placed within a
passively magnetically shielded volume.  The passive magnetic shield
is generally composed of a multi-layer shield formed from thin shells
of material with high magnetic permeability (mu-metal).  The outer
layers of the shield are normally
cylindrical~\cite{bib:nedm2,bib:nedm3.5} or form the walls of a
magnetically shielded room~\cite{bib:altarev2014,bib:altarev2015}.
The innermost magnetic shield is normally a specially shaped shield,
where the design of the coil in relation to shield is carefully taken
into account to achieve adequate homogeneity
\cite{bib:baker,bib:nedm3,bib:nedm5}.

Mechanical and temperature changes of the passive magnetic
shielding~\cite{bib:voigt,bib:thiel}, and the degaussing
procedure~\cite{bib:thiel,bib:altarev2015,bib:fierlinger2016} (also
known as demagnetization, equilibration, or idealization), affect the
stability of the magnetic field within magnetically shielded rooms.
Active stabilization of the background magnetic field surrounding
magnetically shielded rooms can also improve the internal
stability~\cite{bib:voigt,bib:afach,bib:franke}.  The current supplied
to the $B_0$ coil is generated by an ultra-stable current
source~\cite{bib:brys}. The coil must also be stabilized mechanically
relative to the magnetic shielding.

One additional effect, which is the subject of this paper, relates to
the fact that the $B_0$ coil in most nEDM experiments is magnetically
coupled to the innermost magnetic shield.  If the magnetic properties
of the innermost magnetic shield change as a function of time, it then
results in a source of instability of $B_0$.  In the present work, we
estimate this effect and characterize one possible source of
instability: changes of the magnetic permeability $\mu$ of the
material with temperature.

While the sensitivity of magnetic alloys to temperature variations has
been characterized in the past~\cite{bib:couderchon,bib:kruppvdm}, we
sought to make these measurements in regimes closer to the operating
parameters relevant to nEDM experiments.  For these alloys, it is also
known that the magnetic properties are set during the final annealing
process~\cite{bib:gupta,bib:bozorth,bib:kruppvdm}.  In this spirit we
performed our measurements on ``witness'' cylinders, which are small
open-ended cylinders made of the same material and annealed at the
same time as other larger shields are being annealed.

The paper proceeds in the following fashion:
\begin{itemize}
\item The dependence of the internal field on magnetic permeability of
  the innermost shielding layer for a typical nEDM experiment geometry
  is estimated using a combination of analytical and finite element
  analysis techniques.  This sets a scale for the stability problem.
\item New measurements of the temperature dependence of the magnetic
  permeability are presented.  The measurements were done in two ways
  in order to study a variety of systematic effects that were
  encountered.
\item Finally, the results of the calculations and measurements are
  combined to provide a range of temperature sensitivities that takes
  into account sample-to-sample and measurement-to-measurement
  variations.
\end{itemize}

\section{Sensitivity of Internally Generated Field to Permeability of the Shield $B_0(\mu)$\label{sec:calculation}}

The presence of a coil inside the innermost passive shield turns the
shield into a return yoke, and generally results in an increase in the
magnitude of $B_0$.  The ratio of this field inside the coil in the
presence of the magnetic shield to that of the coil in free space is
referred to as the reaction factor $C$, and can be calculated
analytically for spherical and infinite cylindrical
geometries~\cite{bib:bidinostimartin,bib:urankar}.  The key issue of
interest for this work is the dependence of the reaction factor on the
permeability $\mu$ of the innermost shield.  Although this dependence
can be rather weak, the constraints on $B_0$ stability are very
stringent.  As a result, even a small change in the magnetic
properties of the innermost shield can result in an unacceptably large
change in $B_0$.

To illustrate, we consider here the model of a sine-theta surface
current on a sphere of radius $a$, inside a spherical shell of inner
radius $R$, thickness $t$, and linear permeability $\mu$.  The uniform
internal field generated by this ideal spherical coil is augmented by
the reaction factor in the presence of the shield, but is otherwise
left undistorted.  The general reaction factor for this model is given
by Eq.~(38) in Ref.~\cite{bib:bidinostimartin}.  In the high-$\mu$
limit, with $t\ll R$, the reaction factor can be approximated as
\begin{equation}
C 
 \simeq 1+ \frac{1}{2}\, \left( \frac{a}{R} \right)^{3} \left( 1- \frac{3}{2} \, \frac{R}{t} \, \frac{\muo}{\mu} \right) \, ,
 \label{Csphere}
\end{equation}
which highlights the dependence of $B_0$ on the relative permeability
$\mu_r=\mu/\muo$ of the shield.

\begin{figure}[h!]
\begin{center}
   \includegraphics[width=0.7\textwidth]{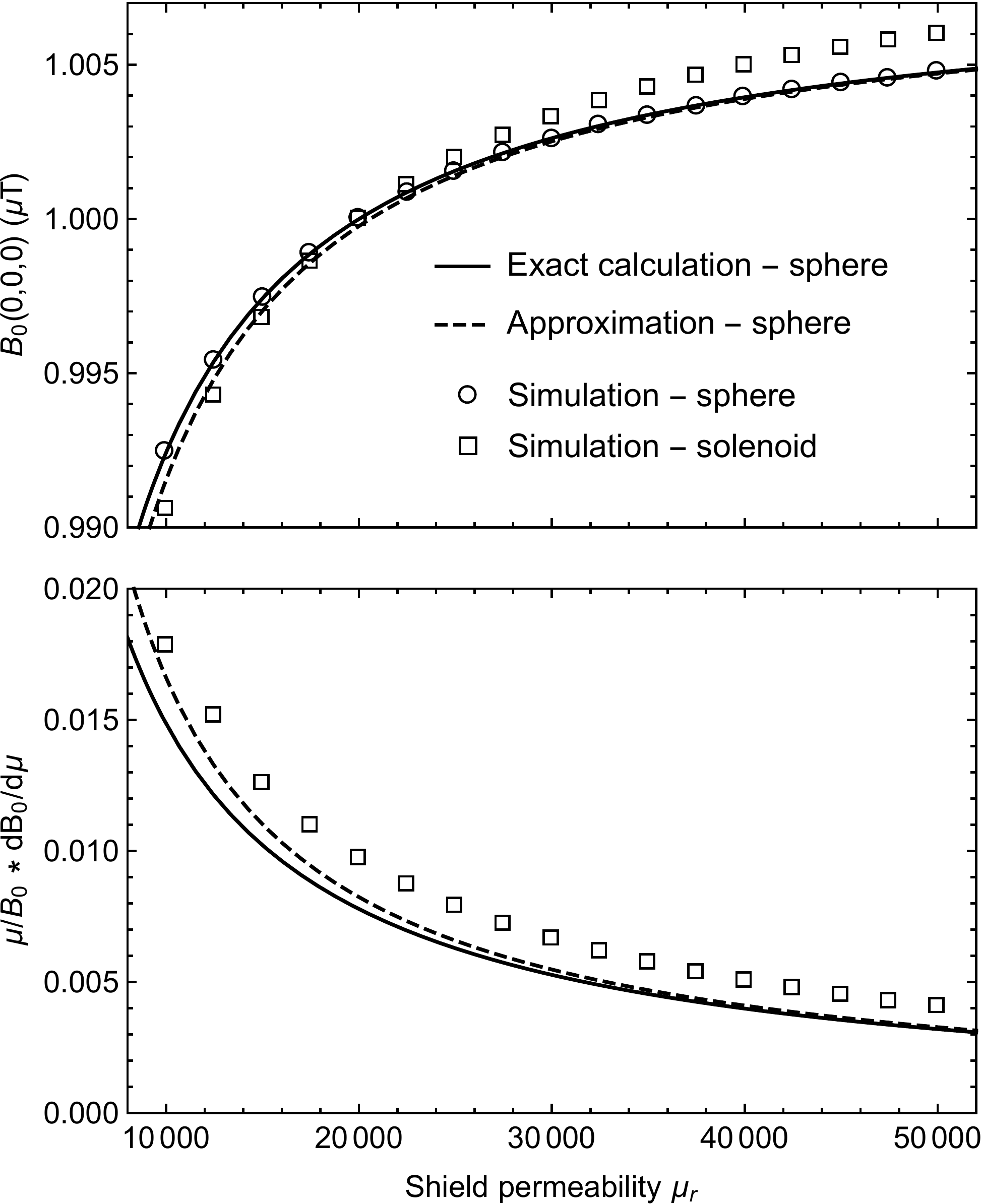}
    \caption{Upper: Magnetic field at the coil center as a function of
      magnetic permeability of the surrounding magnetic shield for a
      geometry similar to the ILL nEDM experiment as discussed in the
      text.  Lower: $\frac{\mu}{B_0}\frac{dB_0}{d\mu}$
      vs.~permeability.  The solid curve is the exact calculation for
      the ideal spherical coil and shield from
      Ref.~\cite{bib:bidinostimartin}; the dashed curve is the
      approximation of Eq.~\ref{Csphere}. The circles and squares are
      the FEMM-based simulations for the spherical and solenoidal
      geometries with discrete currents.  Since the spherical
      simulation was in agreement with the calculation, it is omitted
      from the lower graph.  For the exact calculation and the two
      simulations, currents were chosen to give $B_0=1~\mu$T at
      $\mu_r=20,000$.}
    \label{fig:Magnetic_Field}
    \end{center}
\end{figure}

Fig.~\ref{fig:Magnetic_Field} (upper) shows a plot of $B_0$ versus
$\mu_r$ for coil and shield dimensions similar to the ILL nEDM
experiment~\cite{bib:baker,bib:knecht}: $a=0.53$~m, $R= 0.57$~m, and
$t=1.5$~mm.  In addition to analytic calculations, we also include the
results of two axially symmetric simulations conducted using
FEMM~\cite{bib:femm} to assess the effects of geometry and
discretization of the surface current.  The differences are small,
suggesting that the ideal spherical model of
Ref.~\cite{bib:bidinostimartin} and the high-$\mu$ approximation of
Eq.~\ref{Csphere} provide valuable insight for the design and analysis
of shield-coupled coils.


In the first simulation, the same spherical geometry was used as for
the analytic calculations.  However, the surface current was
discretized to 50 individual current loops, inscribed onto a sphere,
and equally spaced vertically (i.e.~a discrete sine-theta coil).  A
square wire profile of side length 1~mm was used.  As shown in
Fig.~\ref{fig:Magnetic_Field}, this simulation gave excellent
agreement with the analytic calculations.  In the second simulation, a
solenoid coil and cylindrical shield (length/radius~=~2) were used
with the same dimensions as above.  Similarly, the coil was modelled
as 50 evenly spaced current loops, with the distance from an end loop
to the inner face of the shield endcap being half the inter-loop
spacing.  In the limit of tight-packing (i.e., a continuous surface
current) and infinite $\mu$, the image currents in the end caps of the
shield act as an infinite series of current loops, giving the ideal
uniform field of an infinitely long
solenoid~\cite{bib:lambert,bib:sumner}.  As shown in
Fig.~\ref{fig:Magnetic_Field}, the result is similar to the spherical
case, with differences of order one part per thousand and a somewhat
steeper slope of $B_0(\mu_r)$.

Fig.~\ref{fig:Magnetic_Field} (lower) shows the normalized slope
$\frac{\mu}{B_0}\frac{dB_0}{d\mu}$ of the curves from
Fig.~\ref{fig:Magnetic_Field} (upper).  In ancillary measurements of
shielding factors (discussed briefly in
Section~\ref{sec:previousmeasurement}), we found $\mu_r=20,000$ to
offer a reasonable description of the quasistatic shielding factor of
our shield.  Using this value as the magnetic permeability of our
shield material, Fig.~\ref{fig:Magnetic_Field} (lower) shows that
$\frac{\mu}{B_0}\frac{dB_0}{d\mu}$ varies by about 20\% (from 0.008 to
0.01) for the spherical vs.~solenoidal geometries.  We adopt the value
$\frac{\mu}{B_0}\frac{dB_0}{d\mu}=0.01$ as an estimate of this slope
in our discussions in Section~\ref{sec:relationship}, acknowledging
that the value depends on the coil and shield design.

For a high-$\mu$ innermost shield, the magnetic field lines emanating
from the coil all return through the shield.  This principle can be
used to estimate the magnetic field $B_m$ inside the shield material,
and in our studies gave good agreement with FEA-based simulations.
For the solenoidal geometry previously described and used for the
calculations in Fig.~\ref{fig:Magnetic_Field}, $B_m$ is largest in the
side walls of the solenoidal flux return, attaining a maximum value of
170~$\mu$T.  If we assume $\mu_r$=20,000, the $H_m$ field is
0.007~A/m.  Typically the shield is degaussed (idealized) with the
internal coil energized.  After degaussing, $B_m$ must be
approximately the same, since essentially all flux returns through the
shield.  However, the $H_m$ field may become significantly smaller
because after degaussing, it must fall on the ideal magnetization
curve in $B_m-H_m$ space.  (For a discussion of the ideal
magnetization curve, we refer the reader to Ref.~\cite{bib:bozorth}.)
In principle, the $H_m$ field could be reduced by an order of
magnitude or more, depending on the steepness of the ideal
magnetization curve near the origin.  Thus $B_m=170~\mu$T and
$H_m<0.007$~A/m set a scale for the relevant values for nEDM
experiments.  Furthermore, the field in the nEDM measurement volume,
as well as in the magnetic shield, must be stable for periods of
typically hundreds of seconds (corresponding to frequencies
$<0.01$~Hz).  This sets the relevant timescale for magnetic properties
most relevant to nEDM experiments.

\section{Measurements of $\mu(T)$\label{sec:tdep}}

\subsection{Previous Measurements and their Relationship to nEDM Experiments\label{sec:previousmeasurement}}

Previous measurements of the temperature dependence of the magnetic
properties of high-permeability alloys have been summarized in
Refs.~\cite{bib:couderchon,bib:bozorth,bib:pfeifer}.  These
measurements are normally conducted using a sample of the material to
create a toroidal core, where a thin layer of the material is used in
order to avoid eddy-current and skin-depth
effects~\cite{bib:pfeifer,bib:kruppvdm}.  A value of $\mu$ is
determined by dividing the amplitude of the sensed $B_m$-field by the
amplitude of the driving AC $H_m$-field (similar to the method
described in Section~\ref{sec:transformer}).  Normally the frequency
of the $H_m$-field is 50 or 60~Hz.  The value of $\mu$ is then quoted
either at or near its maximum attainable value by adjusting the
amplitude of $H_m$.  Depending on the details of the $B_m-H_m$ curve
for the material in question, this normally means that $\mu$ is quoted
for the amplitude of $H_m$ being at or near the coercivity of the
material~\cite{bib:couderchon,bib:kruppvdm}, resulting in large values
up to $\mu_r=4\times 10^5$.

It is well known that $\mu$ measured in this fashion for toroidal,
thin metal wound cores depends on the annealing process used for the
core.  There is a particularly strong dependence on the take-out or
tempering temperature after the high-temperature portion of the
annealing process has been
completed~\cite{bib:pfeifer,bib:kruppvdm,bib:couderchon}.  Such
studies normally suggest a take-out temperature of 490-500$^\circ$C.
This ensures that the large $\mu_r=4\times 10^{5}$ is furthermore
maximal at room temperature.  Slight variations around room
temperature, and assuming the take-out temperature is not controlled
to better than a degree, imply a scale of possible temperature
variation of $\mu$ of approximately
$\left|\frac{1}{\mu}\frac{d\mu}{dT}\right|\simeq 0.3$-1\%/K at room
temperature~\cite{bib:couderchon,bib:kruppvdm}.

A challenge in applying these results to temperature stability of nEDM
experiments is that, when used as DC magnetic shielding, the
high-permeability alloys are usually operated for significantly
different parameters ($B_m$, $H_m$, and frequencies).

For example, when used in a shielding configuration, the effective
permeability is often measured to be typically $\mu_r=20,000$ rather
than $4\times 10^5$.  This arises in part because $H_m$ is well below
the DC coercivity.  As noted in Section~\ref{sec:calculation}, a more
appropriate $H_m$ for the innermost magnetic shield of an nEDM
experiment is $<0.007$~A/m, whereas the coercivity is
$H_c=0.4$~A/m~\cite{bib:kruppvdm}.  The frequency dependence of the
measurements could also be an issue.  Typically, nEDM experiments are
concerned with slow drifts at $<0.01$~Hz timescales whereas the
previously reported $\mu(T)$ measurements are performed in an AC mode
at 50-60~Hz.

The goal of our experiments was to develop techniques to characterize
the material properties of our own magnetic shields post-annealing, in
regimes more relevant to nEDM experiments.

We created a prototype passive magnetic shield system in support of
this and other precision magnetic field research for the future nEDM
experiment to be conducted at TRIUMF.  The shield system is a
four-layer mu-metal shield formed from nested right-circular
cylindrical shells with endcaps.  The inner radius of the innermost
shield is 18.44~cm, equal to its half-length. The radii and
half-lengths of the progressively larger outer shields increase
geometrically by a factor of 1.27.  Each cylinder has two endcaps
which possess a 7.5~cm diameter central hole.  A stove-pipe of length
5.5~cm is placed on each hole was designed to minimize leakage of
external fields into the progressively shielded inner volumes.  The
design is similar to another smaller prototype shield discussed in
Ref.~\cite{bib:nmorpaper}.  The magnetic shielding factors of each of
the four cylindrical shells, and of various combinations of them, were
measured and found to be consistent with $\mu_r\sim 20,000$.

In our studies of the material properties of these magnetic shields,
two different approaches to measure $\mu(T)$ were pursued.  Both
approaches involved experiments done using witness cylinders made of
the same material and annealed at the same time as the prototype
magnetic shields.  We therefore expect they have the same magnetic
properties as the larger prototype shields, and they have the
advantage of being smaller and easier to perform measurements with.

The two techniques employed to determine $\mu(T)$ were the following:
\begin{enumerate}
\item measuring the low-frequency AC axial magnetic shielding factor
  of the witness cylinder as a function of temperature, and
\item measuring the temperature-dependence of the slope of a minor B-H
  loop, using the witness cylinder as a transformer core, similar to
  previous measurements of the temperature dependence of $\mu$, but
  for parameters closer to those encountered in nEDM experiments.
\end{enumerate}
We now discuss the details and results of each technique.


\subsection{Axial Shielding Factor Measurements\label{sec:axial}}

In these measurements, a witness cylinder was used as a magnetic
shield.  The shield was subjected to a low-frequency AC magnetic field
of $\sim 1$~Hz.  The amplitude of the shielded magnetic field $B_s$
was measured at the center of the witness cylinder using a fluxgate
magnetometer.  Changes in $B_s$ with temperature signify a dependence
of the permeability $\mu$ on temperature.  The relative slope of
$\mu(T)$ can then be calculated using
\begin{equation}
\frac{1}{\mu}\frac{d\mu}{dT}=-\frac{\frac{1}{B_s}\frac{dB_s}{dT}}{\frac{\mu}{B_s}\frac{dB_s}{d\mu}}.
\label{eqn:axial}
\end{equation}
The numerator was taken from the measurements described above. The
denominator was taken from finite-element simulations of the shielding
factor for this geometry as a function of $\mu$.

This measurement technique was sufficiently robust to extract the
temperature dependence of the shielding factor with some degree of
certainty.  Possible drifts and temperature depends of the fluxgate
magnetometer offset were mitigated by using an AC magnetic field.  Any
temperature coefficients in the rest of the instrumentation were
controlled by performing the same measurements with a copper
cylindrical shell in place of the mu-metal witness cylinder.

This technique is quite different than the usual transformer core
measurements conducted by other groups.  As shall be described, it
offers an advantage that considerably smaller $B_m$ and $H_m$ fields
can be accessed.  Measuring the temperature dependence of the
shielding factor is also considerably easier than measuring the
temperature dependence of the reaction factor, since the sensitivity
to changes in $\mu(T)$ is considerably larger in magnitude for the
shielding factor case where $\frac{\mu}{B_s}\frac{dB_s}{d\mu}\sim -1$
compared to the reaction factor case where
$\frac{\mu}{B_0}\frac{dB_0}{d\mu}\sim 0.01$.

\subsubsection{Experimental Apparatus for Axial Shielding Factor Measurements}

The witness cylinder was placed within a homogeneous AC magnetic
field.  The field was created within the magnetically shielded volume
of the prototype magnetic shielding system (described previously in
Section~\ref{sec:previousmeasurement}) in order to provide a
controlled magnetic environment.  A short solenoid inside the
shielding system was used to produce the magnetic field.
The solenoid has 14 turns with 2.6~cm spacing between the wires.  The
solenoid was designed so that the field produced by the solenoid plus
innermost shield approximates that of an infinite solenoid.  The
magnetic field generated by the solenoid was typically 1~$\mu$T in
amplitude.  The solenoid current was varied sinusoidally at typically
1~Hz.

\begin{figure}
  \begin{center}
    \includegraphics[width=0.7\textwidth]{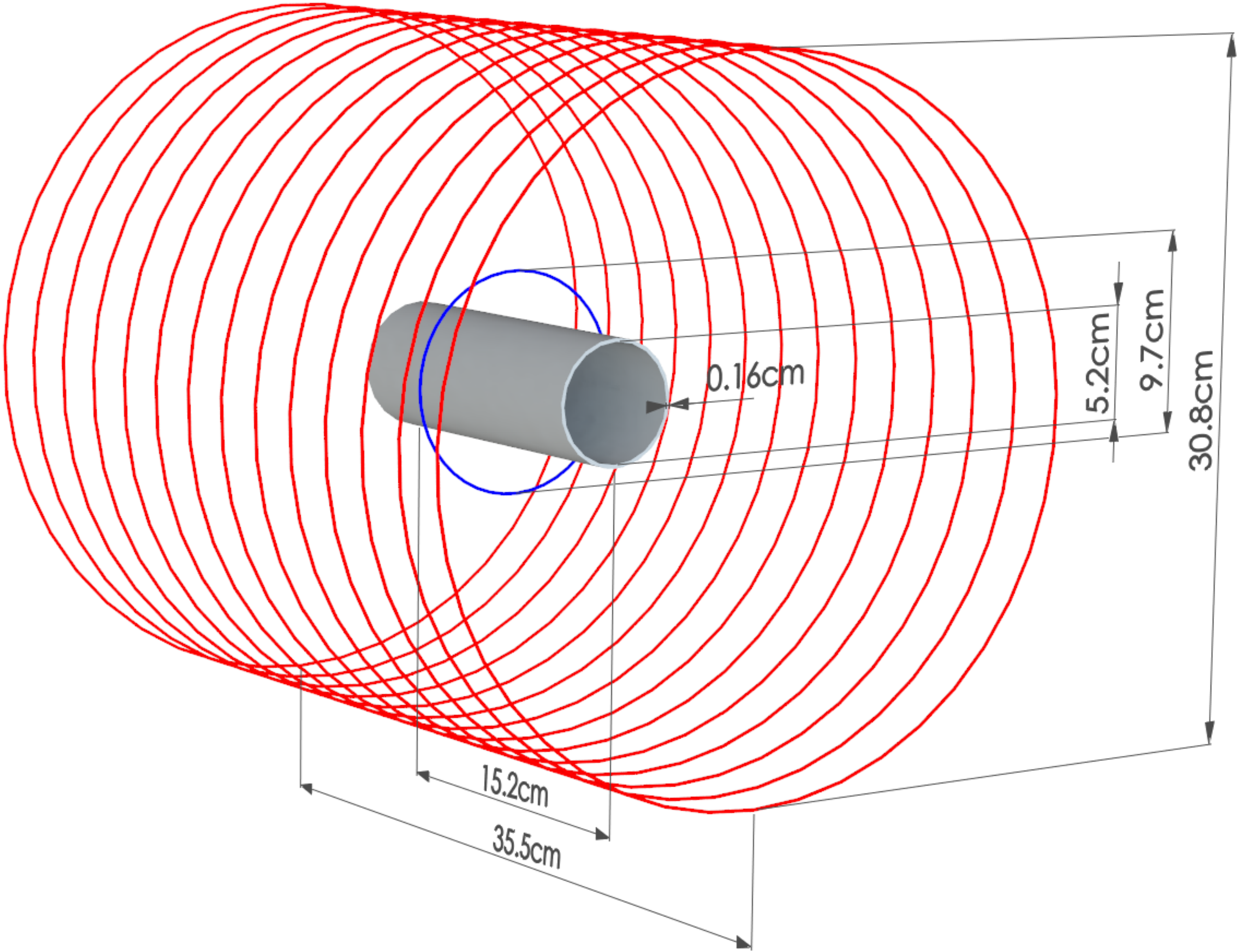}
    \caption{(color online) Axial shielding factor measurement
      setup. The witness cylinder with an inner diameter of 5.2~cm and
      a length of 15.2~cm is placed inside a solenoid (shown in red)
      with a diameter of 30.8~cm and a length of 35.5~cm, containing
      14 turns.  The thickness of the witness cylinder is
      $1/16''=0.16$~cm.  The loop coil (shown in blue) is mechanically
      coupled to the witness cylinder and has a diameter of 9.7~cm.}
    \label{fig:geometry}
  \end{center}
\end{figure}

The witness cylinder was placed into this magnetic field generation
system as shown schematically in Fig.~\ref{fig:geometry}.  The
cylinder was held in place by a wooden stand.

A Bartington fluxgate magnetometer Mag-03IEL70~\cite{bib:bartman} (low
noise) measured the axial magnetic field at the center of the witness
cylinder.  The fluxgate was a ``flying lead'' model, meaning that each
axis was available on the end of a short electrical lead, separable
from the other axes.  One flying lead was placed in the center of the
witness cylinder, the axis of the fluxgate being aligned with that of
the witness cylinder.  The fluxgate was held in place rigidly by a
plastic mounting fixture, which was itself rigidly mounted to the
witness cylinder.

To increase the resolution of the measured signal from the fluxgate, a
Bartington Signal Conditioning Unit (SCU) was used with a low-pass
filter set to typically 10-100~Hz and a gain set to typically $>50$.
The signal from the SCU was demodulated by an SR830 lock-in
amplifier~\cite{bib:lockin} providing the in-phase and out-of-phase
components of the signal.  The sinusoidal output of the lock-in
amplifier reference output itself was normally used to drive the
solenoid generating the magnetic field.  The time constant on the
lock-in was typically set to 3 seconds with 12~dB/oct rolloff.

As shall be described in Section~\ref{sec:axialsyst}, a concern in the
measurement was changes in the field measured by the fluxgate that
could arise due to motion of the system components, or other
temperature dependences.  This could generate a false slope with
temperature that might incorrectly be interpreted as a change in the
magnetic properties of the witness cylinder.

To address possible motion of the witness cylinder with respect to the
field generation system, another coil (the loop coil, also shown in
Fig.~\ref{fig:geometry}) was wound on a plastic holder mounted rigidly
to the witness cylinder.  The coil was one loop of copper wire with a
diameter of 9.7~cm.  Plastic set screws in the holder fixed the loop
coil to be coaxial with the witness cylinder.

Systematic differences in the results from the two coils (the
solenoidal coil, and the loop coil) were used to search for motion
artifacts.  As well, some differences could arise due to the different
magnetic field produced by each coil, and so such measurements could
reveal a dependence on the profile of the applied magnetic field.
This is described further in Section~\ref{sec:axialsyst}.

The temperature of the witness cylinder was measured by attaching four
thermocouples at different points along the outside of the cylinder.
This allowed us to observe the temperature gradient along the witness
cylinder.  To reduce any potential magnetic contamination, T-type
thermocouples were used, which have copper and constantan conductors.
(K-type thermocouples are magnetic.)

Thermocouple readings were recorded by a National Instruments NI-9211
temperature input module.  The magnetic field (signified by the
lock-in amplifier readout) and the temperature were recorded at a rate
of 0.2~Hz.

Temperature variations in the experiment were driven by ambient
temperature changes in the room, although forced air and other
techniques were also tested.  These are described further in
Section~\ref{sec:axialsyst}.

\subsubsection{Data and Interpretation\label{sec:axialsyst}}

\begin{figure}
  \begin{center}
    \includegraphics[width=\textwidth]{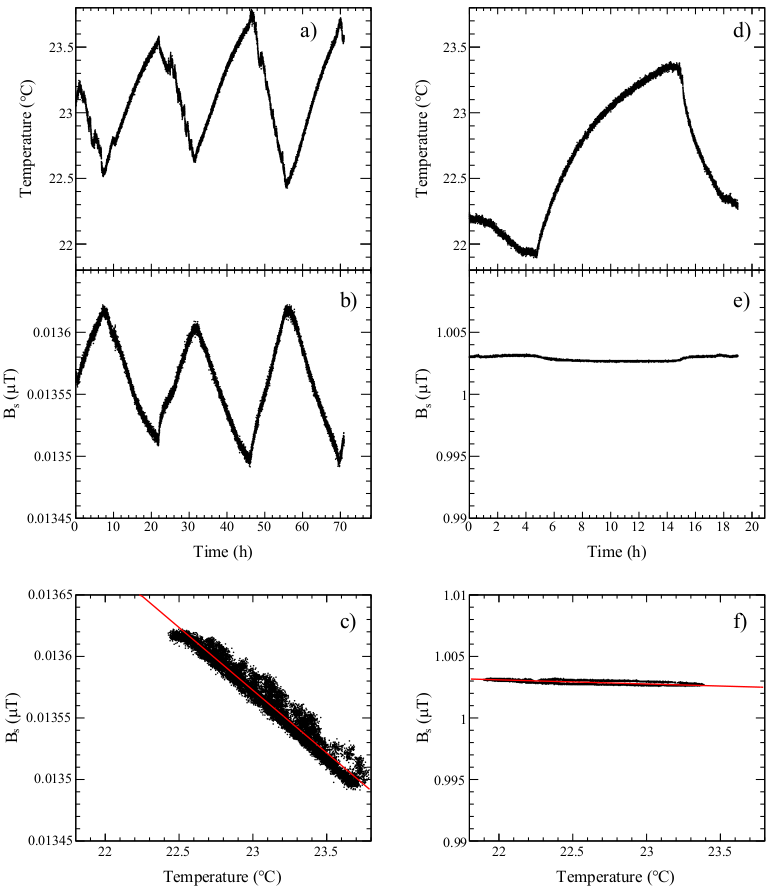}
    \caption{Ambient temperature and shielded magnetic field
      amplitude, measured over a 70 hour period. (a) temperature of
      the witness cylinder as a function of time.  (b) magnetic field
      amplitude measured by fluxgate at center of witness cylinder
      vs.~time.  (c) magnetic field vs.~temperature with linear fit to
      data giving $\frac{1}{B_s}\frac{dB_s}{dT}=-0.75\%$/K (evaluated
      at 23$^\circ$C).  In panels (d), (e), and (f), the same
      quantities are shown for a 20-hour run with a copper cylinder in
      place of the witness cylinder with the linear fit giving
      $\frac{1}{B_s}\frac{dB_s}{dT}=-0.03\%$/K.}
    \label{fig:B_vs_Temp}
  \end{center}
\end{figure}

An example of the typical data acquired is shown in
Fig.~\ref{fig:B_vs_Temp}.  For these data, the field applied by the
solenoid coil was 1~$\mu$T in amplitude, at a frequency of 1~Hz.
Fig.~\ref{fig:B_vs_Temp}(a) shows the temperature of the witness
cylinder over a 70-hr measurement.  The temperature changes of 1.4~K
are caused by diurnal variations in the laboratory.  The shielded
magnetic field amplitude $B_s$ within the witness cylinder is
anti-correlated with the temperature trend as shown in
Fig.~\ref{fig:B_vs_Temp}(b).  Here, $B_s$ is the sum in quadrature of
the amplitudes of the in-phase and out-of-phase components (most of
the signal is in phase).  The magnetic field is interpreted to depend
on temperature, and the two quantities are graphed as a function of
one another in Fig.~\ref{fig:B_vs_Temp}(c).  The slope in
Fig.~\ref{fig:B_vs_Temp}(c) has been calculated using a linear fit to
the data.  The relative slope at 23$^\circ$C was found to be
$\frac{1}{B_s}\frac{dB_s}{dT}=-0.75\%$/K.

Figs.~\ref{fig:B_vs_Temp}(d), (e), and (f) show the same measurement
with essentially the same settings, when the mu-metal witness cylinder
is replaced by a copper cylinder.  A similar relative vertical scale
has been used in Figs.~\ref{fig:B_vs_Temp}(e) and (f) as
Figs.~\ref{fig:B_vs_Temp}(b) and (c).  This helps to emphasize the
considerably smaller relative slope derived from panel (f) compared to
panel (c).  A variety of measurements of this sort were carried out
multiple times for different parameters such as coil current.  Running
the coil at the same current tests for effects due to heating of the
coil, whereas running the coil at a current which equalizes the
fluxgate signal to its value when the mu-metal witness cylinder is
present tests for possible effects related to the fluxgate.  For all
measurements the temperature dependence of the demodulated magnetic
signal was $<0.1$\%/K, giving confidence that unknown systematic
effects contribute below this level.

Some deviations from the linear variation of $B_s$ with $T$ can be
seen in the data, particularly in Figs.~\ref{fig:B_vs_Temp}(a), (b),
and (c).  For example, when the temperature changes rapidly, the
magnetic field takes some time to respond, resulting in a slope in
$B_s-T$ space that is temporarily different than when the temperature
is slowly varying.  This is typical of the data that we acquired, that
the data would generally follow a straight line if the temperature
followed a slow and smooth dependence with time, but the data would
not be linear if the temperature varied rapidly or non-monotonically
with time.  We also tried other methods of temperature control, such
as forced air, liquid flowing through tubing, and thermo-electric
coolers.  The diurnal cycle driven by the building's air conditioning
system gave the most stable method of control and the most
reproducible results for temperature slopes.

As mentioned earlier, data were acquired for both the solenoid coil
and the loop coil.  A summary of the data is provided in
Table~\ref{tab:axial}.  Repeated measurements of temperature slopes
using the loop coil fell in the range
0.4\%/K~$<\vert\frac{1}{B_s}\frac{dB_s}{dT}\vert<$~1.5\%/K.  Similar
measurements for the solenoidal coil yielded
0.3\%/K~$<\vert\frac{1}{B_s}\frac{dB_s}{dT}\vert<$~0.8\%/K.

\begin{table}
\begin{center}
\begin{tabular}{cccc}\hline
Trial & $\frac{1}{B_s}\frac{dB_s}{dT}$ & Coil \\
\#    & (\%/K) & type \\\hline
 1 & -0.32 & solenoid \\
 2 & -0.30 & solenoid \\
 3 & -0.33 & solenoid \\
 4 & -1.53 & loop \\
 5 & -0.42 & loop \\
 6 & -1.30 & loop \\
 7 & -0.74 & solenoid \\
 8 & -1.05 & loop \\
 9 & -0.73 & solenoid \\
10 & -1.23 & loop \\
11 & -0.75 & solenoid \\
12 & -1.12 & loop \\\hline
\end{tabular}
\caption{Summary of data acquired for the AC axial shielding factor
  measurements, in chronological order.  Data with an applied field of
  $\sim 1-6\mu T$ and a measurement frequency of 1~Hz are included.
  Data which used daily fluctuations of the temperature from
  21-24$^\circ$C over a 10-80 hour period are included.  Other data
  acquired for systematic studies are not included in the
  table.\label{tab:axial}}
\end{center}
\end{table}

In general, the slopes measured with the loop coil were larger than
for the solenoidal coil.  This is particularly evident for
measurements 6-12, which were acquired daily over the course of a few
weeks alternating between excitation coils but all used the same
witness cylinder and otherwise without disturbing the measurement
apparatus.  A partial explanation of this difference is offered by the
field profile generated by each coil, and its interaction with the
witness cylinder.  This is addressed further in
Section~\ref{sec:axialsims}.

The other difference between the loop coil and the solenoidal coil was
that the loop coil was rigidly mounted to the witness cylinder,
reducing the possibility of artifacts from relative motion.  Given
that this did not reduce the range of the measured temperature slopes
we conclude that relative motion was well controlled in both cases.

Several other possible systematic effects were considered, all of
which were found to give uncertainties on the measured slopes
$<0.1\%$/K.  These included: thermal expansion of components including
the witness cylinder itself, temperature variations of the magnetic
shielding system within which the experiments were conducted,
degaussing of the witness cylinder, and temperature slopes of various
components e.g. the fluxgate magnetometer and the lock-in amplifier.

As mentioned earlier in reference to Fig.~\ref{fig:B_vs_Temp}(d), (e),
and (f), the stability of the system was also tested by replacing the
mu-metal witness cylinder with a copper cylinder and in all cases
temperature slopes $<0.1$\%/K were measured, giving confidence that
other unknown systematic effects contribute below this level.

Based on the systematic effects that we studied, we conclude that they
do not explain the ranges of values measured for
$\frac{1}{B_s}\frac{dB_s}{dT}$.  We suspect that the range measured is
either some yet uncharacterized systematic effect, or a complicated
property of the material.  We use this range to set a limit on the
slope of $\mu(T)$

\subsubsection{Geometry correction and determination of $\mu(T)$\label{sec:axialsims}}

To relate the data on $B_s(T)$ to $\mu(T)$, the shielding factor of
the witness cylinder as a function of $\mu$ must be known.  Finite
element simulations in FEMM and OPERA were performed to determine this
factor.  The simulations are also useful for determining the effective
values of $B_m$ and $H_m$ in the material, which will be useful to
compare to the case for typical nEDM experiments when the innermost
shield is used as a flux return.

For closed objects, such as spherical
shells~\cite{bib:bidinostimartin,bib:urankar}, the shielding factor
approaches infinity as $\mu \rightarrow \infty$, and
$\vert\frac{\mu}{B_s}\frac{dB_s}{d\mu}\vert\rightarrow 1$.  Because
the witness cylinders are open ended, the shielding factor
asymptotically approaches a constant rather than infinity in the
high-$\mu$ limit, and as a result
$\vert\frac{\mu}{B_s}\frac{dB_s}{d\mu}\vert<1$ here.  From the
simulations the ratio $\frac{\mu}{B_s}\frac{dB_s}{d\mu}$ was
calculated.  A linear model of the material was used where
$\bold{B_m}=\mu\bold{H_m}$ with $\mu$ constant.

The simulations differed slightly in their results, dependent on
whether OPERA or FEMM was used, and whether the solenoidal coil or
loop coil were used.  Based on the simulations, the result is
$\vert\frac{\mu}{B_s}\frac{dB_s}{d\mu}\vert=0.42-0.50$ for the
solenoidal coil, with the lower value being given by FEMM and the
upper value being given by a 3D OPERA simulation, for identical
geometries.  This is somewhat lower than the value suggested by
Ref.~\cite{bib:paperno-open-ended} with fits to simulations performed
in OPERA, which we estimate to be 0.6.  We adopt our value since it is
difficult to determine precisely from
Ref.~\cite{bib:paperno-open-ended}.  For the loop coil, we determine
$\vert\frac{\mu}{B_s}\frac{dB_s}{d\mu}\vert=0.56-0.65$, the range
being given again by a difference between FEMM and OPERA.

Combining the measurement and the simulations, the temperature
dependence of the effective $\mu$ (at $\mu_r=20,000$ which is
consistent with our measurements) can be calculated by
equation~(\ref{eqn:axial}).  The results of the simulations and
measurements are presented in Table~\ref{tab:axialsummary}.  Combining
the loop coil and solenoidal coil results, we find
0.6\%/K~$<\frac{1}{\mu}\frac{d\mu}{dT}<2.7\%$/K to represent the full
range for the possible temperature slope of $\mu$ that observed in
these measurements.

\begin{table}
\begin{center}
\begin{tabular}{|c|c|c|c|}
\hline 
  & $\vert \frac{\mu}{B_s}\frac{dB_s}{d\mu}\vert$ & $\vert \frac{1}{B_s} \frac{dB_s}{dT}\vert$~(\%/K) & $\frac{1}{\mu}\frac{d\mu}{dT}$~(\%/K) \\ 
 & (simulated) & (measured) & (extracted) \\
\hline 
Solenoidal Coil & 0.42-0.50 & 0.3-0.8 & 0.6-1.9 \\ 
\hline 
Loop Coil & 0.56-0.65 & 0.4-1.5 & 0.6-2.7 \\ 
\hline 
\end{tabular} 
\caption{Summary of OPERA and FEMM simulations and shielding factor
  measurements, resulting in extracted temperature slopes of $\mu$.}
\label{tab:axialsummary}
\end{center}

\end{table}

As stated earlier, the simulations also provided a way to determine
the typical $B_m$ and $H_m$ internal to the material of the witness
cylinder.  According to the simulations, the $B_m$ amplitude was
typically 100~$\mu$T and the $H_m$ amplitude was typically 0.004~A/m.
These are comparable to the values normally encountered in nEDM
experiments, recalling from Section~\ref{sec:calculation} that
$H_m<0.007$~A/m for the innermost magnetic shield of an nEDM
experiment.  A caveat is that these measurements were typically
conducted using AC fields at 1~Hz, as opposed to the DC fields
normally used in nEDM experiments.


\subsection{Transformer Core Measurements}
\label{sec:transformer}

An alternative technique similar to the standard method of magnetic
materials characterization via magnetic induction was also used to
measure changes in $\mu$.  In this measurement technique, the witness
cylinder was used as the core of a transformer.  Two coils (primary
and secondary) were wound on the witness cylinder using multistranded
20-gauge copper wire.  The windings were made as tight as possible,
but not so tight as to potentially stress the material.  The windings
were not potted in place.  Three witness cylinders were tested.  Data
were acquired using different numbers of turns on both the primary and
secondary coils (from 6 to 48 on the primary, and from 7 to 24 on the
secondary).

\begin{figure}[h!]
  \begin{center}
    \includegraphics[height=3cm]{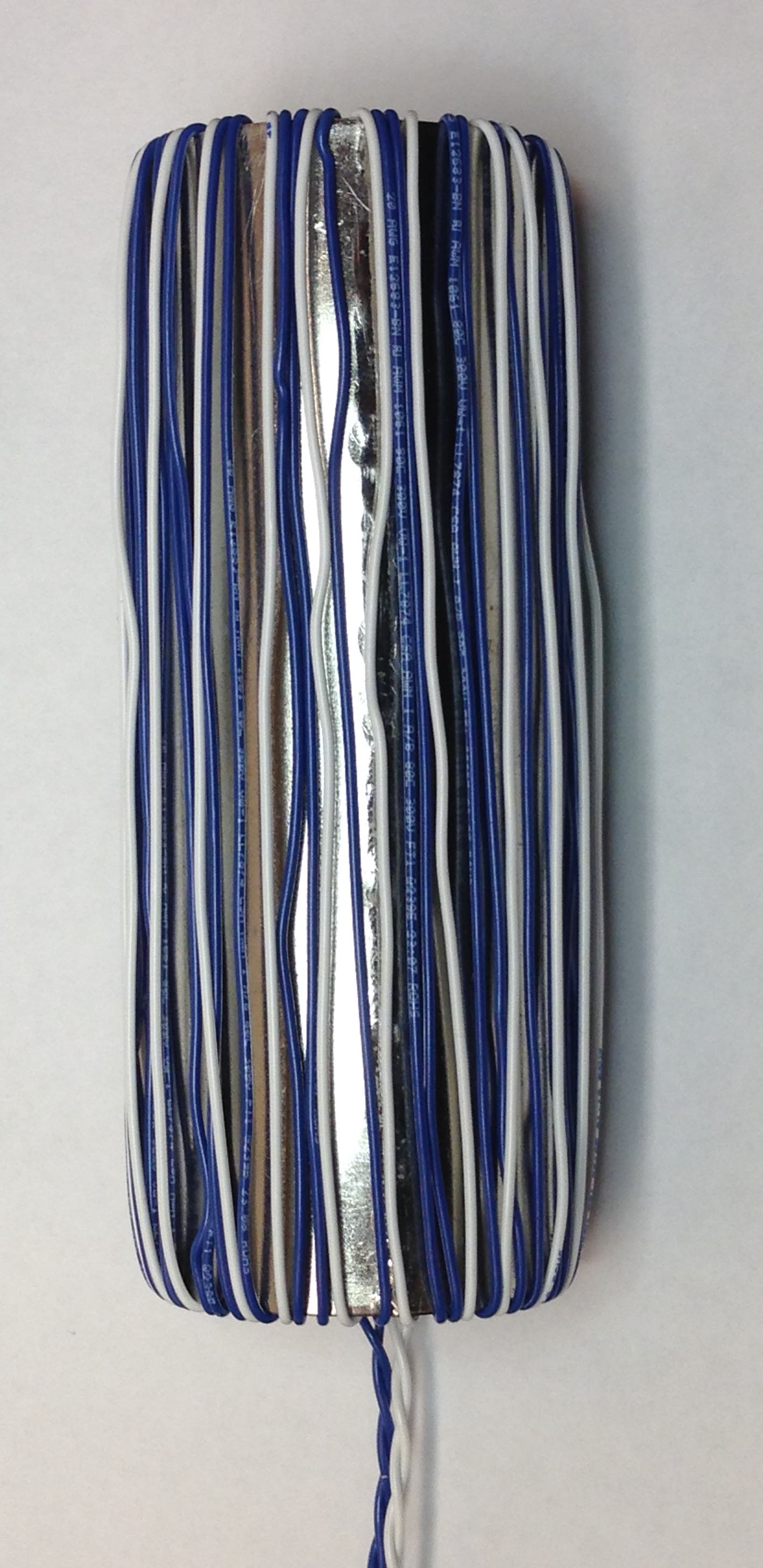}\hskip1cm
    \includegraphics[height=3cm]{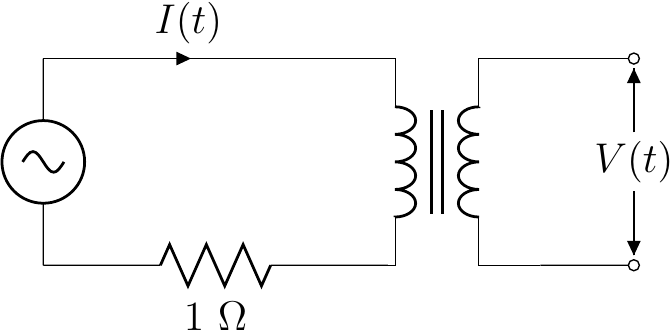}
    \caption{Photograph of a witness cylinder showing transformer
      windings (left) and schematic of the transformer measurement
      (right).  The primary coil was driven by the sine-out of an
      SR830 lock-in amplifier, which was also used to demodulate
      induced voltage $V(t)$ in the secondary coil.  The driving
      current $I(t)$ was sensed by measuring the voltage across a
      stable 1~$\Omega$ resistor.}
    \label{fig:transformer}
  \end{center}
\end{figure}

Fig.~\ref{fig:transformer} shows a picture of one of the witness
cylinders, wound as described.  It also shows a schematic diagram of
the measurement setup, which we now use to describe the measurement
principle.

The primary coil generated an AC magnetic field as a function of time
$H(t)$, while the secondary coil was used to measure the emf induced
by the time-varying magnetic flux proportional to $dB(t)/dt$.  To a
good approximation
\begin{equation}
H_m(t)=\frac{N_pI(t)}{2\pi R}
\end{equation}
where $N_p$ is the number of turns in the primary, $I(t)$ is the
current in the primary, and $R$ is the radius of the witness cylinder,
and
\begin{equation}
\frac{dB_m(t)}{dt}=\dot{B}_m(t)=\frac{V(t)}{b\ell}
\label{eqn:bdot}
\end{equation}
where $V(t)$ is the voltage generated in the secondary, and $b$ and
$\ell$ are the thickness and length of the witness cylinder
respectively.  For a sinusoidal drive current $I(t)$, and under the
assumption that $B_m(t)=\mu H_m(t)$ with $\mu$ being a constant, the
voltage generated in the secondary $V(t)$ should be sinusoidal and out
of phase with the primary current.

The internal oscillator of an SR830 lock-in amplifier was used to
generate $I(t)$.  This was monitored by measuring the voltage across a
1~$\Omega$ resistor with small temperature coefficient in the primary
loop.  The lock-in amplifier was then used to demodulate $V(t)$ into
its in-phase $V_X$ and out-of-phase $V_Y$ components (or equivalently
$\dot{B}_m(t)$ being demodulated into $\dot{B}_{m,X}$ and
$\dot{B}_{m,Y}$, as in equation~(\ref{eqn:bdot})).  The experiment was
done at 1~Hz with $H_m(t)$ as small as possible, typically 0.1~A/m in
amplitude, to measure the slope of the minor $B_m-H_m$ loops near the
origin of the $B_m-H_m$ space.

The temperature of the core was measured continuously using the same
thermocouple arrangement described previously.  Measurements of $V_Y$
as a function of temperature would then signify a change in $\mu$ with
temperature.  In general, we used ambient temperature variations for
the measurements, similar to the procedure used for our axial
shielding factor measurements.

\begin{figure}[h!]
  \begin{center}
    \includegraphics[width=\textwidth]{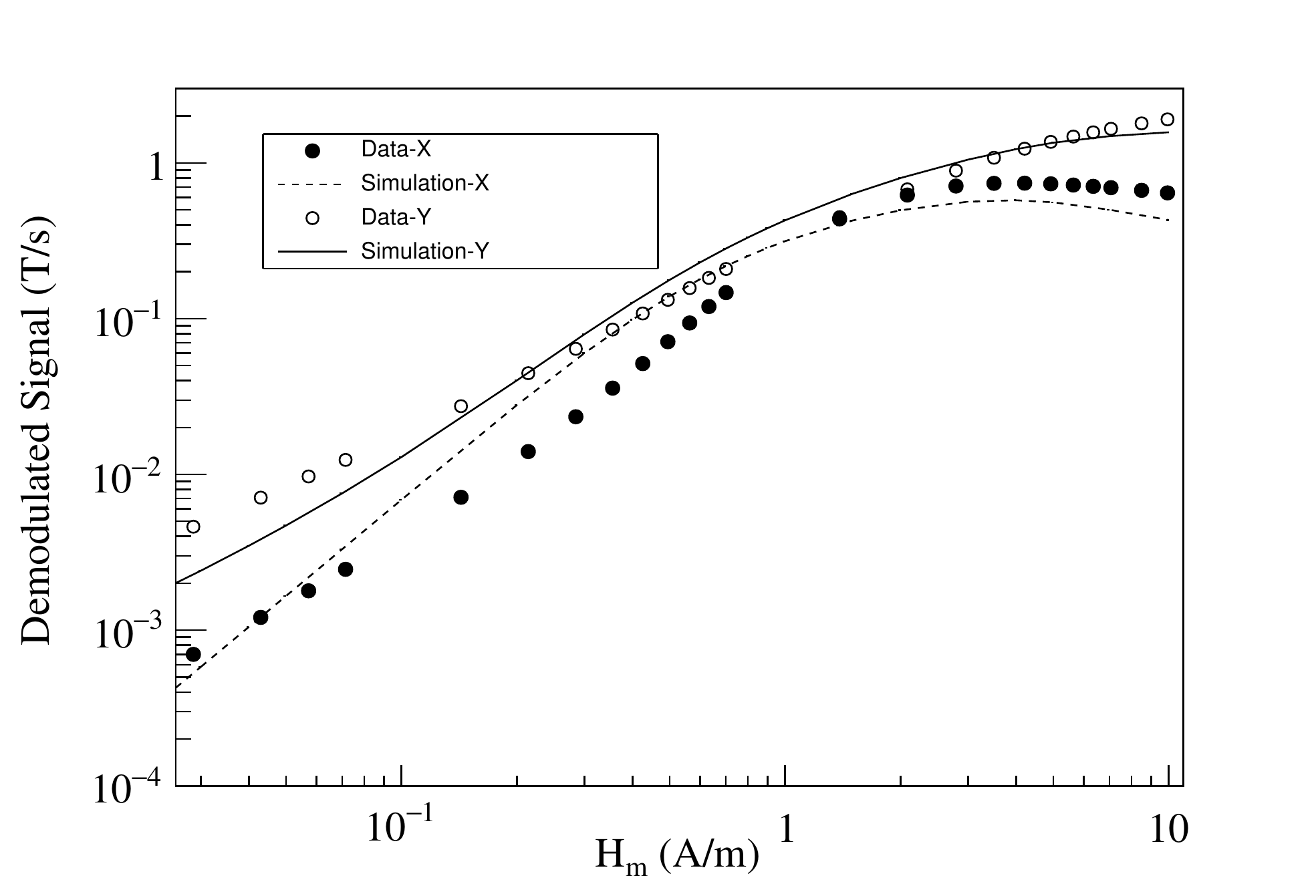}
    \caption{$\dot{B}_{m,X}$ and $\dot{B}_{m,Y}$ as a function of
      amplitude of the applied $H_m$ field at 1~Hz.  Points show the
      acquired data.  Curves display the simulation based on the model
      described in the text.}
    \label{fig:data_and_simulation}
  \end{center}
\end{figure}

The naive expectation is that the out-of-phase $V_Y$ component should
signify a non-zero $\mu$, and the in-phase $V_X$ component should be
zero.  In practice, due to a combination of saturation, hysteresis,
eddy-current losses, and skin-depth effects, the $V_X$ component is
nonzero.  It was found experimentally that keeping the amplitude of
$H_m(t)$ small compared to the apparent coercivity ($\sim 3$~A/m for
the 0.16~cm thick material at 1~Hz frequencies) ensured that the $V_Y$
component was larger than the $V_X$ component.  This is displayed
graphically in Fig.~\ref{fig:data_and_simulation}, where the
dependence of $\dot{B}_{m,Y}$ and $\dot{B}_{m,X}$ on the amplitude of
the applied $H_m(t)$ is displayed, for a driving frequency of 1~Hz.
Clearly the value of $\dot{B}_{m,X}$ can be considerable compared to
$\dot{B}_{m,Y}$, for larger $H_m$ amplitudes near the coercivity.  At
larger amplitudes, the material goes into saturation.  Both
$\dot{B}_{m,Y}$ and $\dot{B}_{m,X}$ eventually decrease as expected at
amplitudes much greater than the coercivity.

To understand the behavior in Fig.~\ref{fig:data_and_simulation}, a
theoretical model of the hysteresis based on the work of
Jiles~\cite{bib:jiles} was used.  The model contains a number of
adjustable parameters.  We adjusted the parameters based on our
measurements of $B_m-H_m$ loops including the initial magnetization
curve.  These measurements were performed separately from our lock-in
amplifier measurements, using an arbitrary function generator and a
digital oscilloscope to acquire them.  The measurements were done at
frequencies from 0.01 to 10~Hz.  It was found that the frequency
dependence predicted by Ref.~\cite{bib:jiles} gave relatively good
agreement with the measured $B_m-H_m$ loops once the five original
(Jiles-Atherton~\cite{bib:jiles-atherton}) parameters were tuned.

For the parameters of the (static) Jiles-Atherton model, we used
$B_s=0.45$~T, $a=3.75$~A/m, $k=2.4$~A/m, $\alpha=2\times 10^{-6}$,
$c=0.05$, which were tuned to our $B_m-H_m$ curve measurements.  For
classical losses, we used the parameters $\rho=5.7\times
10^{-7}~\Omega\cdot$m, $d=1.6$~mm (the thickness of the material), and
$\beta=6$ (geometry factor).  These parameters were not tuned, but
taken from data.  For anomalous losses we used the parameters
$w=0.005$~m and $H_0=0.0075$~A/m, which we also did not tune, instead
relying on the tuning performed in Ref.~\cite{bib:jiles}.

These parameters were then used to model the measurement presented in
Fig.~\ref{fig:data_and_simulation}, including the lock-in amplifier
function.  As shown in Fig.~\ref{fig:data_and_simulation}, trends in
the measurements and simulations are fairly consistent.  The sign of
$\dot{B}_{m,X}$ relative to $\dot{B}_{m,Y}$ is also correctly
predicted by the model (we have adjusted them both to be positive, for
graphing purposes).  We expect that with further tuning of the model,
even better agreement could be achieved.

The model of Ref.~\cite{bib:jiles} makes no prediction of the
temperature dependence of the parameters.  Ideally, the temperature
dependence of $\dot{B}_{m,Y}$ and $\dot{B}_{m,X}$ under various
conditions could be used to map out the temperature dependence of the
parameters.  However, this is beyond the scope of the present work.

We make the simplifying assumption that temperature dependence of
$\dot{B}_{m,Y}$ may be approximately interpreted as the temperature
dependence of a single parameter $\mu$, i.e. that
\begin{equation}
\frac{1}{\dot{B}_{m,Y}}\frac{d\dot{B}_{m,Y}}{dT}=\frac{1}{\mu}\frac{d\mu}{dT}.
\end{equation}
This is justified in part by our selection of measurement parameters
(the amplitude of $H_m=0.1$~A/m and a measurement frequency of 1~Hz)
which ensure that $\dot{B}_{m,Y}$ dominates over $\dot{B}_{m,X}$.

We assign no additional systematic error for this simplification, and
all our results are subject to this caveat.  We comment further that
in our measurements of the axial shielding factor (presented in
Section~\ref{sec:axial}), the same caveat exists.  In that case the
in-phase component dominates the demodulated fluxgate signal.  In a
sense, measuring $\mu(T)$ itself is always an approximation, because
it is actually the parameters of minor loops in a hysteresis curve
which are measured.  In reality, our results may be interpreted as a
measure of the temperature-dependence of the slopes of minor loops
driven by the stated $H_m$.

Measurements of $\frac{1}{\dot{B}_{m,Y}}\frac{d\dot{B}_{m,Y}}{dT}$ as
a function of $T$ were made.  In general, the data mimicked the
behavior of the axial shielding factor measurements, giving a similar
level of linearity with temperature as the data displayed in
Fig.~\ref{fig:B_vs_Temp}.  Other similar behaviors to those
measurements were also observed, for example: (a) when the temperature
slope changed sign, $\dot{B}_{m,Y}$ would temporarily give a different
slope with temperature, (b) the measured value of
$\frac{1}{\dot{B}_{m,Y}}\frac{d\dot{B}_{m,Y}}{dT}$ depended on a
variety of factors, most notably a dependence on which of the three
witness cylinders was used for the measurement, and on differences
between subsequent measurements using the same cylinder.

\begin{table}
\begin{center}
\begin{tabular}{cccc}\hline
Trial & $\frac{1}{\dot{B}_{m,Y}}\frac{d\dot{B}_{m,Y}}{dT}$ & core \\
\#    & (\%/K) & used \\\hline
 1 & 0.15 & $\alpha$ \\
 2 & 0.03 & $\alpha$ \\
 3 & 0.04 & $\alpha$ \\
 4 & 0.06 & $\alpha$ \\
 5 & 1.07 & $\beta$  \\
 6 & 0.93 & $\beta$  \\
 7 & 0.88 & $\beta$  \\
 8 & 0.88 & $\beta$  \\
 9 & 0.09 & $\alpha$ \\
10 & 1.23 & $\beta$  \\
11 & 2.15 & $\beta$  \\
12 & 1.85 & $\beta$  \\
13 & 1.20 & $\beta$  \\
14 & 0.77 & $\gamma$ \\\hline
\end{tabular}
\caption{Summary of data acquired for the transformer core
  measurements.  Three different witness cylinders, arbitrarily
  labeled $\alpha$, $\beta$, and $\gamma$, were used for the
  measurements.  A 1~Hz excitation frequency was used with amplitudes
  for $H_m$ ranging from 0.1 to 0.3~A/m.  Fluctuations in the
  temperature ranged from 21-24$^\circ$C and measurement times over a
  10-80 hour period are included.  Other data acquired for systematic
  studies are not included in the table.\label{tab:transformer}}
\end{center}
\end{table}

Table~\ref{tab:transformer} summarizes our measurements of the
relative slope $\frac{1}{\dot{B}_{m,Y}}\frac{d\dot{B}_{m,Y}}{dT}$ for
a variety of trials, witness cylinders, and numbers of windings.  The
data show a full range of $0.03-2.15$\%/K for
$\frac{1}{\mu}\frac{d\mu}{dT}=\frac{1}{\dot{B}_{m,Y}}\frac{d\dot{B}_{m,Y}}{dT}$,
again naively assuming the material to be linear as discussed above.
The sign of the slope of $\mu(T)$ was the same as the axial shielding
factor technique.

A dominant source of variation between results in this method arose
from properties inherent to each witness cylinder.  One of the
cylinders (referred to as $\beta$ in Table~\ref{tab:transformer}) gave
temperature slopes consistently larger
$\frac{1}{\mu}\frac{d\mu}{dT}\sim 0.88-2.15$\%/K than the other two
$\frac{1}{\mu}\frac{d\mu}{dT}\sim 0.03-0.77$\%/K (referred to as
$\alpha$ and $\gamma$, with some evidence that $\gamma$ had a larger
slope than $\alpha$).  We expect this indicates some difference in the
annealing process or subsequent treatment of the cylinders, although
to our knowledge the treatment was controlled the same as for all
three cylinders.  Since our goal is to provide input to future EDM
experiments on the likely scale of the temperature dependence of $\mu$
that they can expect, we phrase our result as a range covering all
these results.

Detailed measurements of the effect of degaussing were conducted for
this geometry.  The ability to degauss led us ultimately to select a
larger number of primary turns (48) so that we could fully saturate
the core using only the lock-in amplifier reference output as a
current source.  A computer program was used to control the lock-in
amplifier in order to implement degaussing.  A sine wave with the
measurement frequency (typically 1~Hz) was applied at the maximum
lock-in output power.  Over the course of several thousand
oscillations, the amplitude was decreased linearly to the measurement
amplitude ($\sim 0.1$~A/m).  After degaussing with parameters
consistent with the recommendations of
Refs.~\cite{bib:thiel,bib:altarev2015}, the measured temperature
slopes were consistent with our previous measurements where no
degaussing was done.

Other systematic errors found to contribute at the $<0.1\%$/K level
were: motion of the primary and secondary windings, stability of the
lock-in amplifier and its current source, and stability of background
noise sources.

To summarize, the dominant systematic effects arose due to different
similarly prepared cores giving different results, and due to
variations in the measured slopes in multiple measurements on the same
core.  The second of these is essentially the same error encountered
in our axial shielding factor measurements.  We expect it has the same
source; it is possibly a property of the material, or an additional
unknown systematic uncertainty.

\section{Relationship to nEDM experiments\label{sec:relationship}}

Neutron EDM experiments are typically designed with the DC coil being
magnetically coupled to the innermost magnetic shield.  As discussed
in Section~\ref{sec:calculation}, if the magnetic permeability of the
shield changes, this results in a change in the field in the
measurement region by an amount
$\frac{\mu}{B_0}\frac{dB_0}{d\mu}=0.01$.

The temperature dependence of $\mu$ has been constrained by two
different techniques using open-ended mu-metal witness cylinders
annealed at the same time as our prototype magnetic shields.  We
summarize the overall result as
0.0\%/K~$<\frac{1}{\mu}\frac{d\mu}{dT}<$~2.7\%/K, where the range is
driven in part by material properties of the different mu-metal
cylinders, and in part by day-to-day fluctuations in the temperature
slopes.

We note the following caveats in relating this measurement to nEDM
experiments:
\begin{itemize}
\item Although the measurement techniques rely on considerably larger
  frequencies and different $H_m$-fields than those relevant to
  typical nEDM experiments, we think it reasonable to assume the
  temperature dependence of the effective permeability should be of
  similar scale.  For frequency, both techniques typically used a 1~Hz
  AC field, whereas for nEDM experiments the field is DC and stable at
  the 0.01~Hz level.  Furthermore, in one measurement technique the
  amplitude of $H_m$ was $\sim 0.004$~A/m and in the other was $\sim
  0.1$~A/m.  For nEDM experiments $H_m<0.007$~A/m and is DC.
\item Both measurement techniques extract an effective $\mu$ that
  describes the slope of minor loops in $B_m-H_m$ space.  A more
  correct treatment would include a more comprehensive accounting of
  hysteresis in the material, which is beyond the scope of this work.
\end{itemize}

Assuming our measurement of
0.0\%/K~$<\frac{1}{\mu}\frac{d\mu}{dT}<2.7$\%/K and the generic EDM
experiment sensitivity of $\frac{\mu}{B_0}\frac{dB_0}{d\mu}=0.01$
results in a temperature dependence of the magnetic field in a typical
nEDM experiment of $\frac{dB_0}{dT}=0-270$~pT/K.  To achieve a goal of
$\sim 1$~pT stability in the internal field for nEDM experiments, the
temperature of the innermost magnetic shield in the nEDM experiment
should then be controlled to the $<0.004$~K level if the worst-case
dependence is to be taken into account.  This represents a potentially
challenging design constraint for future nEDM experiments.

As noted by others~\cite{bib:cpviolwithoutstrangeness}, the use of
self-shielded coils to reduce the coupling of the $B_0$ coil to the
innermost magnetic shield is an attractive option for EDM experiments.
The principle of this technique is to have a second coil structure
between the inner coil and the shield, such that the net magnetic
field generated by the two coils is uniform internally but greatly
reduced externally.  For a perfect self-shielded coil, the field at
the position of the magnetic shield would be zero, resulting in
perfect decoupling, which is to say a reaction factor that is
identically unity.  For ideal geometries, such as spherical
coils~\cite{bib:brown, bib:wheeler,bib:purcell} or infinitely long
sine-phi coils~\cite{bib:bethBNL,bib:bethUSpatent,bib:bidinosti}, the
functional form of the inner and outer current distributions are the
same, albeit with appropriately scaled magnitudes and opposite sign.
More sophisticated analytical and numerical methods have been used
extensively in NMR and MRI to design self-shielded
gradient~\cite{bib:turner,bib:hidalgo},
shim~\cite{bib:brideson,bib:forbes}, and transmit
coils~\cite{bib:bidinosti,bib:kuzmin}, and should be of value in the
context of nEDM experiments, as well.  We are also pursuing novel
techniques for the design of self-shielded coils of any arbitrary
field profile and geometric shape~\cite{bib:crawford}.

\section{Conclusion}

In the axial shielding factor measurement, we found
0.6\%/K~$<\frac{1}{\mu}\frac{d\mu}{dT}<2.7\%$/K, with the measurement
being conducted with a typical $H_m$-amplitude of 0.004~A/m and at a
frequency of 1~Hz.  In the transformer core case, we found
0.0\%/K~$<\frac{1}{\mu}\frac{d\mu}{dT}<2.2\%$/K, with the measurement
being conducted with a typical $H_m$-amplitude of 0.1~A/m and at a
frequency of 1~Hz.

The primary caveat to these measurements is that both measurements
(transformer core and axial shielding factor) do not truly measure
$\mu$.  Rather they measure observables related to the slope of minor
hysteresis loops in $B_m-H_m$ space.  They would be more appropriately
described by a hysteresis model like that of Jiles~\cite{bib:jiles},
but to extract the temperature dependence of all the parameters of the
model is beyond the scope of this work.  Instead we acknowledge this
fact and relate the temperature dependence of the effective $\mu$
measured by each experiment.

We think it is interesting and useful information that the two
experiments measure the same scale and sign of the temperature
dependence of their respective effective $\mu$'s.  This is a principal
contribution of this work.

In future work, we plan to measure $B_0(T)$ directly for nEDM-like
geometries using precision atomic magnetometers.  We anticipate based
on the present work that self-shielded coil geometries will achieve
the best time and temperature stability.

\section{Acknowledgments}

We thank D.~Ostapchuk from The University of Winnipeg for technical
support.  We gratefully acknowledge the support of the Natural
Sciences and Engineering Research Council Canada, the Canada
Foundation for Innovation, and the Canada Research Chairs program.

\section*{References}

\bibliography{mybibfile}

\begin{thebibliography}{00}

\bibitem{bib:nedm2} A. P. Serebrov {\it et al.}, JETP Lett. {\bf 99}, 4
  (2014).

\bibitem{bib:nedm2.5} A. P. Serebrov {\it et al.}, Phys. Procedia {\bf
  17}, 251 (2011).

\bibitem{bib:nedm3} K. Kirch, AIP Conf. Proc. {\bf 1560}, 90 (2013).

\bibitem{bib:nedm3.5} C. A. Baker, {\it et al.}, Phys. Procedia {\bf
  17}, 159 (2011).

\bibitem{bib:nedm5} I. Altarev, {\it et al.}, Nuovo Cim. C {\bf
  35}, 122 (2012).

\bibitem{bib:nedm6} R. Golub and S. K. Lamoreaux, Phys. Rept.  {\bf
  237}, 1 (1994).

\bibitem{bib:nedm6.5} T. M. Ito (for the nEDM Collaboration),
  J. Phys. Conf. Ser. {\bf 69} 012037, 2007.

\bibitem{bib:nedmtriumf} R.~Picker (for the TRIUMF Japan-Canada UCN
  Collaboration), in the proceedings of MENU2016, July 25-30, 2016,
  Kyoto, Japan, arXiv:1612.00875 [physics.ins-det].

\bibitem{bib:baker} C. A. Baker, {\it et al.}, Phys. Rev. Lett. {\bf
  97}, 131801 (2006).

\bibitem{bib:pendlebury} J. M. Pendlebury {\it et al.}, Phys. Rev. D
  {\bf 92}, 092003 (2015).

\bibitem{bib:brys} T. Bry\'s, {\it et al.}, Nucl. Instrum. Meth. A
  {\bf 554}, 527 (2005).

\bibitem{bib:afach} S. Afach, {\it et al.}, J. Appl. Phys. {\bf 116},
  084510 (2014).

\bibitem{bib:fierlingerroom} I. Altarev, {\it et al.}
  Rev. Sci. Instrum. {\bf 85}, 075106 (2014).

\bibitem{bib:sturmthesis} M. Sturm, Masterarbeit, T.U. M\"unchen
  (2013).

\bibitem{bib:patton} B. Patton, E. Zhivun, D. C. Hovde, and D. Budker,
  Phys. Rev. Lett. {\bf 113}, 013001 (2014).


\bibitem{bib:altarev2014} I. Altarev {\it et al.},
  Rev. Sci. Instrum. {\bf 85}, 075106 (2014).


\bibitem{bib:altarev2015} I. Altarev {\it et al.}, J. Appl. Phys. {\bf 117}, 233903 (2015).



\bibitem{bib:voigt} J. Voigt {\it et al.}, Metrol. Meas. Syst. {\bf 20}, 239 (2013).

\bibitem{bib:thiel} F. Thiel {\it et al.}, Rev. Sci. Instrum. {\bf 78}, 035106 (2007).

\bibitem{bib:fierlinger2016} Z. Sun {\it et al.}, J. Appl. Phys. {\bf
  119}, 193902 (2016).

\bibitem{bib:franke} B. Franke, PhD Thesis, ETH Z\"urich (2013).

\bibitem{bib:couderchon} G. Couderchon, J. F. Tiers,
  J. Magn. Magn. Mat. {\bf 26}, 196 (1982).

\bibitem{bib:kruppvdm} Krupp VDM Magnifer 7904, Material Data Sheet
  No.~9004, Aug. 2000, Krupp VDM GmbH, Postfach 18 20, D-58778
  Werdohl, Germany.

\bibitem{bib:gupta} K. Gupta, K. K. Raina, S. K. Sinha, J. Alloys
  Compd. {\bf 429}, 357 (2007).

\bibitem{bib:bozorth} R. M. Bozorth, {\it Ferromagnetism} (IEEE Press,
  Piscataway, NJ, 1993).

\bibitem{bib:bidinostimartin} C. P. Bidinosti, J. W. Martin, AIP Advances
  {\bf 4}, 047135 (2014).

\bibitem{bib:urankar} L. Urankar, R. Oppelt, IEEE
  Trans. Biomed. Eng. {\bf 43}, 697 (1996).

\bibitem{bib:knecht} A. Knecht, PhD Thesis, U. Z\"urich (2009).

\bibitem{bib:femm} Finite Element Method Magnetics FEMM version 4.2,
  available from {\tt http://www.femm.info}.


\bibitem{bib:lambert} R.H.\ Lambert and C.\ Uphoff, Rev. Sci. Instrum. {\bf 46},  337 (1975).

\bibitem{bib:sumner} T.J.\ Sumner,  J. Phys. D: Appl. Phys. {\bf 20} 692 (1987).






\bibitem{bib:pfeifer} F. Pfeifer and C. Radeloff,
  J. Magn. Magn. Mat. {\bf 19}, 190 (1980).

\bibitem{bib:nmorpaper} J. W. Martin {\it et al.},
  Nucl. Instrum. Meth. A {\bf 778}, 61 (2015).

\bibitem{bib:bartman} Bartington Instruments Ltd., 10 Thorney Leys
  Business Park, Witney, Oxon, OX28 4GG, England.

\bibitem{bib:lockin} Stanford Research Systems, 1290-D Reamwood Ave.,
  Sunnyvale, CA 94089.

\bibitem{bib:paperno-open-ended} E. Paperno, IEEE Trans. Magn. {\bf
  35}, 3940 (1999).

\bibitem{bib:jiles} D. C. Jiles, J. Appl. Phys. {\bf 76}, 5849 (1994).

\bibitem{bib:jiles-atherton} D. C. Jiles and D. L. Atherton,
  J. Appl. Phys. {\bf 55}, 2115 (1984); D. C. Jiles and D. L. Atherton,
  J. Magn. Magn. Mat. {\bf 61}, 48 (1986).



\bibitem{bib:cpviolwithoutstrangeness} I. B. Khriplovich,
  S. Lamoreaux, {\it CP violation without strangeness: electric dipole
    moments of particles, atoms, and molecules.} (Springer-Verlag,
  Berlin, 2012).



\bibitem{bib:brown} W. F. Brown Jr. and J. H. Sweer, Rev. Sci. Instrum. {\bf 16},  276 (1945).

\bibitem{bib:wheeler} H. A. Wheeler, Proceedings of the IRE {\bf 46},1595 (1958).  

\bibitem{bib:purcell}E.M.\ Purcell, 
Am. J. Phys. \textbf{57}, 18 (1989); Am. J. Phys. \textbf{58}, 296 (1990). 




\bibitem{bib:bethBNL} R.A.\  Beth, Brookhaven National Laboratory Report 
BNL-10143 (1966). 

\bibitem{bib:bethUSpatent} R.A.\  Beth, 
US Patent 3466499, September 9, 1969.


\bibitem{bib:bidinosti} C.P. Bidinosti, I.S. Kravchuk, and
  M.E. Hayden, J. Magn. Res. {\bf 177}, 31 (2005).
  

\bibitem{bib:turner} R.\ Turner and R.M.\ Bowley, 
J. Phys. E: Sci. Instrum. \textbf{19}, 876 (1986). 

\bibitem{bib:hidalgo} S.S.\ Hidalgo-Tobon, 
Concepts Magn. Reson. \textbf{36A}, 223 (2010).  

\bibitem{bib:brideson} M.A. Brideson, L.K. Forbes, S. Crozier, Concepts Magn. Reson. {\bf 14}, 9 (2002).

\bibitem{bib:forbes} L.K. Forbes and S. Crozier, J. Phys. D: Appl. Phys {\bf 36}, 68 (2003).

\bibitem{bib:kuzmin} V.V.\ Kuzmin  {\it et al.}, J. Magn. Reson. {\bf 256}, 70 (2015).

\bibitem{bib:crawford} C. Crawford, \textit{private communication}.





\end{thebibliography}

\end{document}